\documentstyle[preprint,osa,epsf]{revtex}
\begin{document}

\title{Nonlinearity Management in Dispersion Managed System
\footnote{\bf Accepted to Optics Letters (2001)} }

\author{I.R. Gabitov$^{1}$ and P. M. Lushnikov$^{1,2}$}

\address{$^1$ Theoretical Division, Los Alamos National Laboratory,
  MS-B284, Los Alamos, New Mexico, 87545
  \\
  $^2$ Landau Institute for Theoretical Physics, Kosygin St. 2,
  Moscow, 117334, Russia }


\maketitle

\begin{abstract}
  We propose using a nonlinear phase shift interferometric converter
  (NPSIC) (a new device) for lumped compensation of the nonlinearity in
  optical fibers.  The NPSIC is a nonlinear analog of the Mach-Zehnder
  interferometer and provides a way to control the sign of the nonlinear
  phase shift. We investigate a potential use of NPSIC for
  compensation of the nonlinearity to make a dispersion-managed system
  closer to an ideal linear system.  More importantly NPSIC can be used to
  essentially improve single channel capacity in the nonlinear regime.

\end{abstract}
~~~~~~~ {\it OCIS codes:}  060.2330, 060.5530, 060.4370, 190.5530, 260.2030.
\\


  The recent invention~\cite{80Kogelnik} and
  testing~\cite{kurtzke1,chrapl1,99Gerges,00Mol2} of the dispersion
  management technique demonstrated the effectiveness of this approach
  for high speed communications.  Optical pulse dynamics in fiber
  links with dispersion management are governed by the nonlinear
  Schr\"odinger equation (NLS) with periodic coefficients:

\begin{eqnarray}
i u_z +d(z)    u_{tt} + \sigma(z) |u|^{2} u =0, \label{nls1}
\end{eqnarray}
where $z$ is the propagation distance, $u$ is an optical pulse amplitude, $d(z)\equiv -\frac{1}{2}
\beta_2(z) $, $\beta_2(z)$ is a first order group-velocity dispersion, $\sigma(z)=(2\pi n_2)/\big
(\lambda_0 A_{eff}(z)\big)$ is the nonlinear coefficient, $n_2$ is the nonlinear refractive index,
$\lambda_0=1.55\, \mu m$ is the carrier wavelength, $A_{eff}$ is the effective fiber area that in
general case depends on $z$.

On short scales, the dispersion managed (DM) system is practically linear. Linear transmission in
an optical fiber is limited by a nonlinear distance $z<z_{nl}\equiv (\sigma |u_0|^2)^{-1}$, which
is determined by the Kerr nonlinearity $\sigma$ and the characteristic pulse power $|u_{0}|^2$.
The characteristic power cannot be choosen too small in order to maintain appropriate value of the
signal-to-noise ratio.  It is natural to attempt to extend the scale of the applicability of the
linear regime.  This can be achieved through use of a new optical fiber with lower Kerr
nonlinearity~\cite{Hamaide95,Weins97}. Another obvious approach is ``nonlinearity management'',
which was considered in~\cite{pare1996}. However, the semiconductor material waveguides with
negative Kerr nonlinearity proposed as an element for compensation of nonlinear phase shift are
currently not practical.  In this letter we consider lumped compensation of the nonlinearity, the
analog of lumped compensation of the chromatic dispersion by means of chirped fiber gratings. We
suggest the use of a nonlinear phase shift interferometric converter (NPSIC) as presented in
Fig.1. The NPSIC consists of a silica-based fiber 1, a highly nonlinear fiber 2 (which could be
chalcogenide glass based, having a Kerr coefficient about 400 times or even much more than that
for silica, see e.g. reference~\cite{chal2}), a linear amplifier A with amplitude amplification
coefficient $G$, and two directional couplers 1,2.  It is assumed that an optical terminator is
installed at the end of the fiber 2 after coupler 2 to prevent beam reflection from the end of
fiber 2. The incident light with amplitude $Gu_0$ is split by the directional coupler 1 into two
beams with amplitudes $a_1G u_0$ and $a_2G u_0 e^{i\pi/2}$ ($a_1>0, \ a_2>0$), corresponding to
the fibers 1 and 2, respectively. We assume that total power is conserved: $a_1^2+a_2^2=1$ but a
subsequent consideration can be easily generalized to include the directional coupler's and
fiber's losses. The extra phase $\pi/2$ in the second fiber is due to light spliting in the
directional coupler (see e.g. reference~\cite{kashima1995}). The optical lengths $l_1, \ l_2$ in
fibers 1 and 2 between couplers 1 and 2 are chosen in such a way that they provide a zero phase
difference at the input of coupler 2: $n_{eff, 1}l_1-n_{eff, 2}l_2=0$, where $n_{eff, 1},n_{eff,
2}$ are effective linear refraction indexes in fibers 1,2. We assume that $l_1, l_2$ are small
enough so that we can neglect the influence of dispersion in both fibers and the nonlinear phase
shift in fiber 1. Thus the amplitudes of optical beams at the input of the coupler 2 are given by
$a_1G u_0$ and $a_2G u_0 e^{i \pi/2+i\phi_{nl}}$, where the nonlinear phase shift
$\phi_{nl}=\sigma_{hnl}a_2^2G^2 |u_0|^2 l_2$ and $\sigma_{hnl}$ correspond to the value of
$\sigma$ in the highly nonlinear fiber 2. Assuming that the propagation constants are the same for
symmetric and antisymmetric modes of the coupler, the coupled-wave equation describing mode
evolution in directional couplers can be written as\cite{kashima1995}
\begin{eqnarray}\label{psi1psi2}
(u_1)_z= i\kappa u_2, \quad (u_2)_z = i\kappa u_1,
\end{eqnarray}
where
\begin{eqnarray}\label{z0}
u_1(z_0)=a_1G u_0, \ u_2(z_0)=a_2G u_0 e^{i \pi/2+i\phi_{nl}},
\nonumber\\
u_{1\, out}=u_1(z_{out}),
\end{eqnarray}
$z_0, \, z_{out}$ are the coordinates of coupler 2 input and output respectively; $u_{1\, out}$ is
the optical beam amplitude in fiber 1 at the exit of the coupler 2 and $\kappa$ is a coupling
coefficient which is assumed to be real in the lossless model.  The solution of
$(\ref{psi1psi2}),(\ref{z0})$ shows that NPSIC converts the $u_0$ into the output signal $u_{1\,
out}$ as follows
\begin{equation}\label{psisum1}
u_{1\, out}= Gu_0 \big ( a_1 \cos{\phi_c}-a_2 e^{i\phi_{nl}}\sin{\phi_c} \big ),
\end{equation}
 where $\phi_c=\kappa(z_{out}-z_0).$ We assume that $\phi_{nl}\equiv
\sigma_{hnl}a_2^2G^2 |u_0|^2 l_2\ll 1$ and expand $e^{i\phi_{nl}}$ in the Eq. $(\ref{psisum1})$.
Neglecting $O(\phi_{nl}^2)$ terms results in
\begin{equation}\label{psisum2}
u_{1\, out}= u_{lin}\big ( 1-\frac{i \sigma_{hnl}a_2^3G^2 |u_0|^2 l_2 \sin{\phi_c}}{ a_1
\cos{\phi_c}-a_2 \sin{\phi_c}} \big ),
\end{equation}
where $u_{lin}\equiv Gu_0 (a_1 \cos{\phi_c}-a_2 \sin{\phi_c} )$ is the output amplitude $u_{1\,
out}$ of the linear system.  Thus by changing the NPSIC parameters $a_1, \, \phi_c, \, l, \, G $,
one can control the sign and magnitude of the nonlinear phase shift.  NPSIC can be considered as a
nonlinear version of the Mach-Zehnder interferometer. As a typical example we take $G=\sqrt{50},
\, a_1=1/3, \, \phi_c=0.2$ which results in $u_{1\, out}\simeq u_0\big ( 1.0-i 60. \, \sigma_{hnl}
|u_0|^2 l_2 \big ).$ Thus to completely compensate a nonlinear phase shift $\sigma |u_0|^2 L_1$ in
$L_1=40\, km$ line of a silica fiber we need to use NPSIC with $l_2=1.6\, m$ of a highly nonlinear
fiber provided that $\sigma_{hnl}/\sigma=400$ (for this value of $\sigma_{hnl}$ the nonlinear
absorption is negligible in currently available chalcogenide glasses\cite{chal2}).

This estimate is true if the power $|u|^2$ is constant throughout the propagation. In practice,
power is not constant, due to fiber chromatic dispersion and losses. Therefore, nonlinearity
compensators must be distributed along the fiber with a separation distance less than the
dispersion length $z_{disp}\equiv \tau^2/d$ and loss length $z_{loss}$, where $\tau$ is a typical
pulse width. It is shown below that the effective length of the nonlinearity is increased as
$z_{eff,\, nl}\sim N^{2} z_{nl}$, where $N$ is the number of lumped NPSIC on the dispersion map
period. However, we demonstrate that for short pulses corresponding to strong dispersion
management this approach is efficient only at relatively large value of $N$. Nevertheless, this
method displays good performance in the nonlinear regime, as, for example, inserting only two
compensating elements on the period of the dispersion map could increase the bit rate per channel
by a factor of two.

 Consider DM system with step-wise periodical dispersion variation:
 $d(z)=d_0+\tilde d(z)$, $\tilde d(z)=d_1$ for $0<z+m L<L_1$ (standard monomode
 fiber) and $\tilde d(z)=d_2$ for $L_1<z+m L<L_1+L_2$ (dispersion
 compensating fiber). Here $d_0$ is the path-averaged dispersion,
 $d_1,d_2$ are the amplitudes of dispersion variation subject to a
 condition $d_1L_1+d_2L_2=0$, $L\equiv L_1+L_2$ is a dispersion
 compensation period and $m$ is an arbitrary integer number.
 $\sigma(z)=\sigma_{1}$ for $0<z+n L<L_1$ and $\sigma(z)=\sigma_{2}$
 for $L_1<z+n L<L_1+L_2$ corresponding to standard monomode and dispersion
 compensating fiber. We suppose that NPSIC units are located inside
 and at the ends of a dispersion compensated fiber at points $z_n=m
 L+L_1+ n L_2/N,$ $n=0,\ldots, N,$ $ z_0=mL+L_1<z_1< \ldots <
 z_N=(m+1)L$. At these points the value of $u(t,z)$ experiences jump
 according to the Eq.  $(\ref{psisum2})$:
\begin{equation}\label{nonlinjump1}
u(t,z)\big |_{z=z_{m,n}+0}=(u- i z_{eff, n} \, |u|^2u)\big |_{z=z_{m,n}-0},
\end{equation}
where $z_{m,n}\equiv m L+L_1+ n L_2/N$; $z_{m,n}-0$ and $z_{m,n}+0$ mean the coordinate value just
before and after the jump, respectively; the parameters $a_1, \, \phi_c, \, l, \, G $ are chosen
in such a way in order to to provide $u_{lin}=u(t,z)|_{z=z_{m,n}-0}$ and $\sigma_{hnl}a_2^3G^3 l_2
\sin{\phi_c} \equiv z_{eff, n}\equiv (\sigma_{1}L_1+\sigma_{2}L_2)/N$ for n=1,\ldots, N-1 and
$z_{eff, 0}\equiv z_{eff, N}\equiv (\sigma_{1}L_1+\sigma_{2}L_2)/(2N).$ Here the term
$-(\sigma_{1}L_1+\sigma_{2}L_2)$ provides the compensation of the nonlinear phase shift both in
standard monomode and dispersion compensating fibers.

Assuming that the nonlinearity is small $z_{nl}\gg z_{disp}, \ z_{nl}\gg L$, where $L$ is a
dispersion map period, one can express $u$ in Fourier domain as a product of an exact solution of
linear part of the Eq. $(\ref{nls1})$, corresponding to mean-free dispersion $\tilde d(z)$, on a
slow function $\hat \psi(\omega,z)$: $\hat u(\omega,z) \equiv \hat \psi(\omega,z)
e^{-i\omega^2\int^z_{z_0} \tilde d(z')dz'}$ (see reference~\cite{gabtur1}), where $\hat
u(\omega,z)=\int^\infty_{-\infty}u(t,z)e^{\imath \omega t}dt$. $\hat \psi$ is a slow function of
$z$ on a scale $L$ which allows to integrate the Eq. $(\ref{nls1})$ over the period $L$ neglecting
the slow dependence of $\hat \psi$ on $z$ and we get by taking into account the jumps in
$(\ref{nonlinjump1})$:
\begin{eqnarray} \label{nlsomega2}
i \hat \psi(\omega,(m+1) L)-i\hat \psi(\omega,m L) - L \omega^2 d_0 \hat \psi(\omega,m L)
\nonumber \\
+ \frac{1}{(2\pi)^2}\int  \hat \psi(\omega_1,m L) \hat \psi(\omega_2, m L)\hat \psi^*(\omega_3,m L)
 \\
 \times
K_{tot}(\triangle)\delta(\omega_1+\omega_2-\omega-\omega_3)d\omega_1 d\omega_2d\omega_3=0,
 \nonumber
\end{eqnarray}
where $s\equiv d_1 L_1$, $\triangle \equiv \omega^2_1+\omega^2_2-\omega^2-\omega^2_3$ and
$K_{tot}(\triangle)\equiv K_{1}(\triangle)+K_{2}(\triangle)$. A kernel $K_1(\triangle)$ is equal
to the usual kernel of the path-averaged equation~\cite{gabtur1}:
\begin{eqnarray} \label{K1}
K_1(\triangle)=\sigma_{L}\sin{x}/x, \, x\equiv s\triangle/2, \, \sigma_{L}\equiv
\sigma_{1}L_1+\sigma_{2}L_2,
\end{eqnarray}
while $K_2(\triangle)$ accounts for the jumps in $(\ref{nonlinjump1})$:

\begin{eqnarray} \label{K2}
K_2(\triangle)= -\sigma_L[\sin{x}\, \cot{(x/N)}]/N.
\end{eqnarray}
Suppose that Full Width at Half Maximum (FWHM) $\tau$ of $\psi(0,t)$ is subject to the condition
$s/(2N\tau^2)\ll 1$ then $K_{tot}(\triangle)$ can be rewritten as:

\begin{eqnarray} \label{Ktot2}
K_{tot}(\triangle)=\sigma_L (\sin{x}/x)  \Big[ x^2/(3N^2) +O\Big (x^2 /N^2\Big )\Big].
\end{eqnarray}
 Thus comparing $(\ref{Ktot2})$ with the kernel $(\ref{K1})$ of the
usual path-averaged equation we get an extra small factor $\Big (\frac{s\triangle}{2N}\Big
)^2/3\sim L^2/(N z_{disp})^2$ provided that $N$ is big enough to ensure the condition $\frac{L}{N
z_{disp}}\ll 1.$ For short pulses $N$ should be a big number to make the system close to the
linear one. Figure 2 shows the dependence of the FWHM $\tau_{out}$ obtained after propagation of
the initial zero-chirp Gaussian pulse with $\tau_{ini}=10\, ps$ over a typical transoceanic
distance $10^4\,  km$ on the number of NPSIC units $N+1$. This dependence is obtained by numerical
intergation of the NLS equation~$(\ref{nls1})$. The pulse is lunched to the DM fiber at $z=L_1/2$
with the peak power $|u|^2=2\, mW$.  DM system parameters are $d_0=0, \ d_1=10.0 \, ps^2/km,\
d_2=-101.6\, ps^2/km, L_1=40\, km, \ L_2=-d_1 L_1/d_2, \ \sigma_1=0.0013 \, (km \, mW)^{-1}, \
\sigma_2=0.00405\, (km \, mW)^{-1}$. It is seen that the system can be approximately considered
linear for $N\stackrel{>}{\sim}30$ in which case the nonlinear correction to $\tau_{out}$ is less
$5\%$.

A strong variation of $\tau_{out}$ in Fig. 2 for small $N$ is due to the attraction of the initial
Gaussian pulse to soliton solution $\hat\psi(\omega, mL)\equiv \hat\psi_0(\omega)e^{imL\lambda}$ of
path-averaged Eq. $(\ref{nlsomega2})$. Here $\lambda$ is a soliton propagation constant. We refer
to this soliton solution as the modified DM soliton (MDM soliton) by analogy with the DM soliton,
that corresponds to the usual path-average equation (without nonlinearity management). For
$K_2(\triangle)=0$ we recover the usual DM soliton of the path-averaged equation~\cite{gabtur1}
where the DM soliton width $\tau_{DM}$ depends on $s$ only for $d_0\to 0$ (see e.g.
reference~\cite{lush2000a}). For the above mentioned system parameters $\tau_{DM}\simeq 21 ps.$
Any shorter pulses experience strong distortion because of the nonlinearity for
$K_2(\triangle)\equiv 0.$ Curve 2 in Fig. 3 shows the pulse power distribution after $10^4 km$ of
propagation of the initial Gaussian pulse with $\tau_{ini}=10\, ps$ (curve 1). Using NPSIC units
we can control $K_2(\triangle)$ and thus change $\tau_{MDM}$ of MDM soliton. Curve 3 shows the
pulse power distribution after $10^4\, km$ for a system with two NPSIC for each DM period $L$
located at the ends of the dispersion compensating fiber with
$z_{eff,0}=z_{eff,1}=f(\sigma_{1}L_1+\sigma_{2}L_2)/2$, where f=3/2. One can see that curve 3 is
close both to the initial Gaussian pulse (curve 1) and to the MDM soliton (curve 4). The MDM
soliton was obtained by numerical iteration of the equation $(\ref{nlsomega2})$ with
$\hat\psi(\omega, mL)= \hat\psi_0(\omega)e^{imL\lambda}, \ \lambda =0.00028 \, km^{-1}$. The
numerical iteration scheme is similar to the one used in reference~\cite{lush2001a} for usual
path-averaged equation. By changing the factor $f$ one can control $\tau_{MDM}$ of the MDM
soliton. Note that for $f>1$, which corresponds to a negative average nonlinearity of the total
system, the numerical iteration scheme finally diverges indicating that the MDM soliton is a
rather long-lived quasi-stable structure which however serves as an attractor of the pulse
dynamics on a scale $\sim 10^4 \, km.$

In summary, the use of NPSIC elements would allow one to construct a nearly linear fiber
transmission system. However this system requires many NPSIC elements.  On the other hand the use
of NPSIC elements offers the better advantages to fiber links operating in nonlinear regime. In
particular, only a few elements per dipersion map period could dramatically reduce pulse width and
potentially increase bit-rate.

The authors thanks M. Chertkov and E.A. Kuznetsov for helpful discussions.

The support was provided by the Department of Energy, under contract W-7405-ENG-36.

E-mail addresses: lushnikov@cnls.lanl.gov, ildar@t7.lanl.gov


\newpage

\section*{Figure captions:}

~

\noindent Fig.1. Schematic of the nonlinear phase shift interferometric converter.

\noindent Fig.2. FWHM $\tau_{out}$ after pulse propagation over $10^4$ km versus a number $N+1$ of
NPSIC units.

\noindent Fig.3. Power distributions:  an initial Gaussian pulse (curve 1); result of pulse
propagation over $10^4$ km in a DM system with no NPSIC (curve 2); result of pulse propagation in
a DM system with two NPSIC units for each period $L$ (curve 3); the MDM soliton (curve 4).


\begin{figure*}
\epsfxsize=15.5cm \epsffile{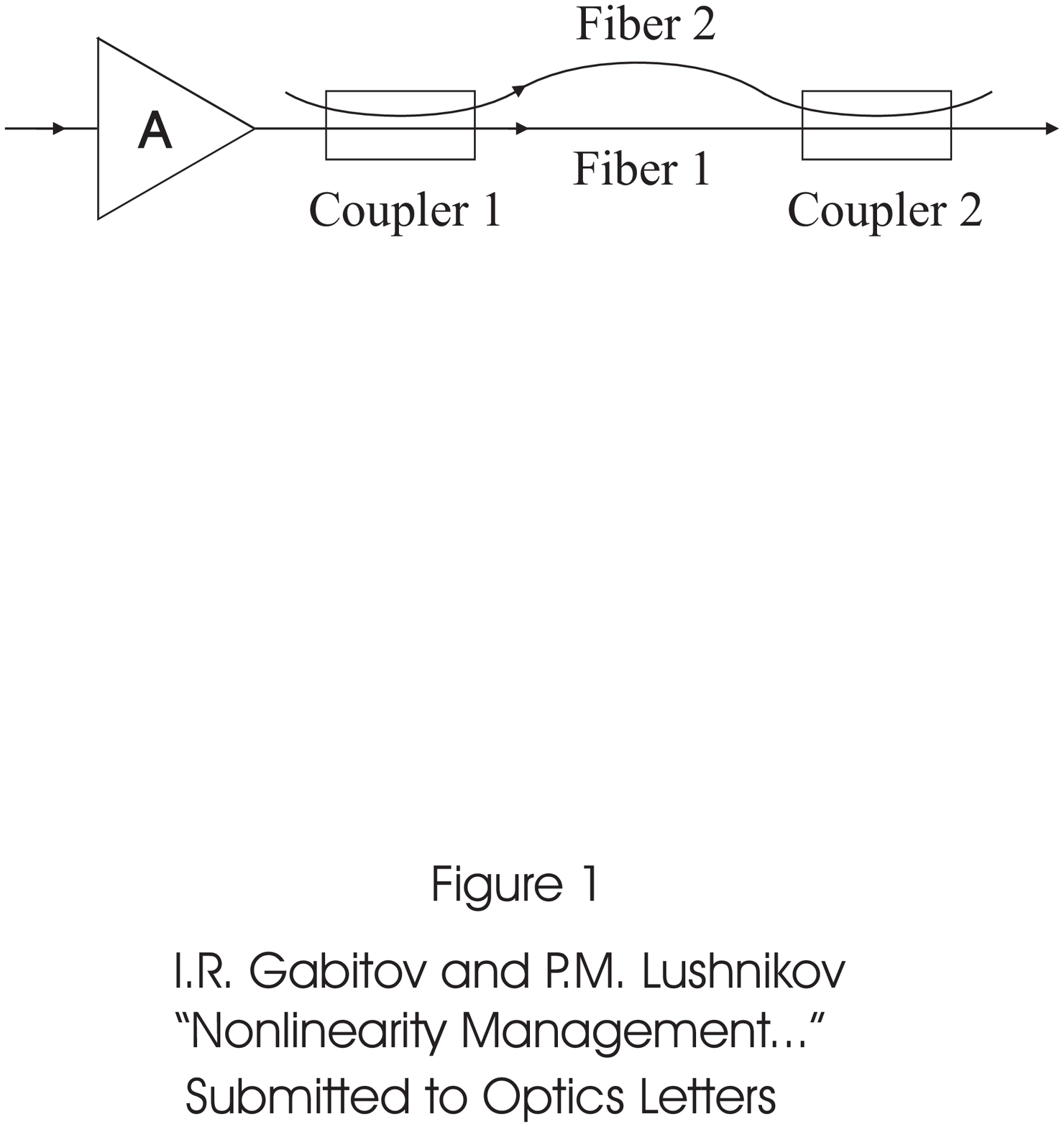}
\end{figure*}

\begin{figure*}
\epsfxsize=15.5cm \epsffile{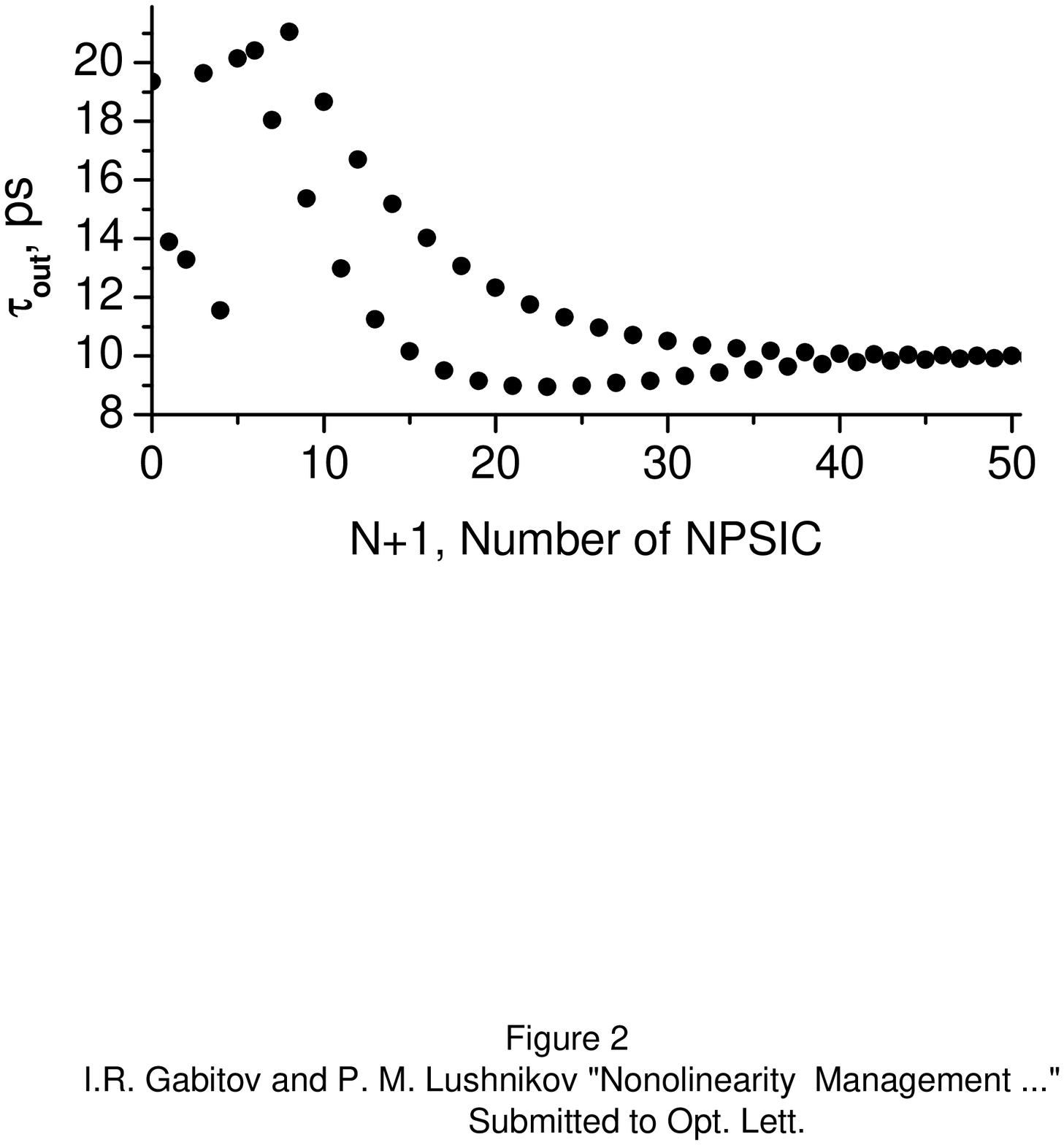}
\end{figure*}

\begin{figure*}
\epsfxsize=15.5cm \epsffile{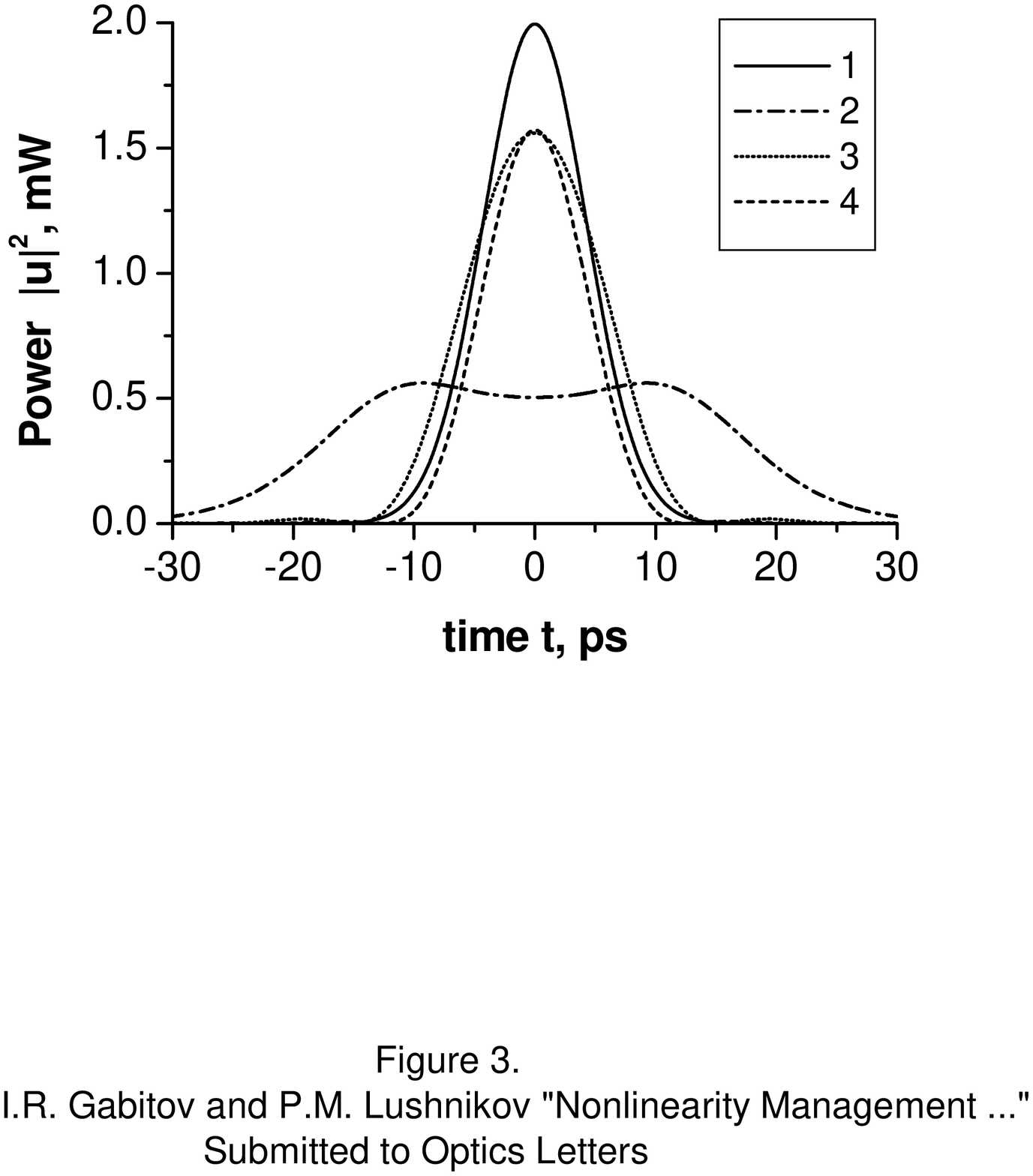}
\end{figure*}




\end{document}